\documentclass[a4paper,fleqn,usenatbib]{mnras}
\usepackage{graphicx}

\usepackage[T1]{fontenc}
\usepackage{ae,aecompl}

\usepackage{graphicx}	
\usepackage{amsmath}	
\usepackage{amssymb}	

\title[3D kinematics of NGC 6362]{3D core kinematics of NGC~6362: central rotation in a 
dynamically evolved globular cluster\thanks{Based on observations collected at the European Organisation 
for Astronomical Research in the Southern Hemisphere under ESO programme 0103.D-0545 (PI: Dalessandro).}}

\author[Emanuele Dalessandro]{
Emanuele Dalessandro,$^{1}$\thanks{E-mail: emanuele.dalessandro@inaf.it}
Silvia Raso,$^{1}$
Sebastian Kamann,$^{2}$
Michele Bellazzini,$^{1}$  \newauthor
Enrico Vesperini,$^{3}$
Andrea Bellini,$^{4}$
and Giacomo Beccari$^{5}$
\\
$^{1}$INAF - Osservatorio di Astrofisica e Scienza dello Spazio di Bologna, Via Gobetti 93/3, Bologna I-40129, Italy\\
$^{2}$Astrophysics Research Institute, Liverpool John Moores University, 146 Brownlow Hill, Liverpool L3 5RF, UK\\
$^{3}$Department of Astronomy, Indiana University, Swain West, 727 E. 3rd Street, IN 47405 Bloomington - USA\\
$^{4}$Space Telescope Science Institute, 3700 San Martin Drive, Baltimore, MD 21218, USA\\
$^{5}$European Souther Observatory, Karl Schwarzschild Strasse 2, D-85748 Garching bei Munchen, Germany
}

\date{Accepted 2021 April 28. Received 2021 April 26; in original form 2021 February 8}

\pubyear{2021}

\begin{document}
\label{firstpage}
\pagerange{\pageref{firstpage}--\pageref{lastpage}}
\maketitle

\begin{abstract}
We present a detailed 3D kinematic analysis of the central regions ($R<30\arcsec$) of the low-mass and dynamically evolved 
galactic globular cluster NGC 6362. The study is based on data obtained with 
ESO-VLT/MUSE used in combination with the adaptive optics module and providing $\sim3000$ line-of-sight 
radial velocities, which have been complemented 
with Hubble Space Telescope proper motions. 
The quality of the data and the number of available radial velocities allowed us 
to detect for the first time a significant rotation signal along the line of sight 
in the cluster core with amplitude of $\sim 1$ km/s and with a peak 
located at only $\sim20\arcsec$ from the cluster center, corresponding to only $\sim10\%$ of the cluster half-light radius.  
This result is further supported by the detection of a central and significant 
tangential anisotropy in the cluster innermost regions. 
This is one of the most central rotation signals ever observed in a globular cluster
to date.
We also explore the rotational properties of the multiple populations hosted by this cluster 
and find that Na-rich stars rotate about two times more rapidly than the Na-poor sub-population thus suggesting 
that the interpretation of the present-day globular cluster properties require 
a multi-component chemo-dynamical approach.
Both the rotation amplitude and peak position would fit qualitatively the theoretical expectations for a system 
that lost a significant fraction of its original mass because of the long-term dynamical evolution and interaction 
with the Galaxy. However, to match the observations more 
quantitatively further theoretical studies to explore the initial  
dynamical properties of the cluster are needed.
\end{abstract}




\section{Introduction}
Globular clusters (GCs) are true touchstones for Astrophysics. Their study can bring crucial information on a variety 
of subjects ranging from stellar evolution \citep{cassisi13} to the initial epochs of star formation in the Universe 
and eventually to the formation 
and mass assembly history of their host galaxies (e.g. \citealt{brodie_strader06,dalessandro12,forbes18,krumholz19}).

Traditionally GCs have been considered as relatively simple spherical, non-rotating and almost completely relaxed systems.
However, observational results obtained in the past few years are demonstrating that they
are much more complex than previously thought. In particular, the classical simplified approach of neglecting rotation in GCs has become untenable from
the observational point of view.
In fact, there is an increasing wealth of observational results suggesting that, when properly surveyed, 
the majority of GCs rotate at some level.
As of today, more than $50\%$ of the sampled GCs show clear signatures of internal rotation 
(e.g., \citealt{and03,bellazzini12,fabricius14,ferraro18,lanzoni18a,lanzoni18b,kamann18a,bianchini18,sollima19}).
Moreover, evidence of rotation has also been reported for intermediate-age clusters \citep{mackey13,kamann18a}, 
young massive clusters \citep{henault12,dalessandro21} and nuclear star clusters \citep{nguyen18,neumayer20} indicating that internal rotation is a common ingredient across dense stellar systems of different 
sizes and ages.
On the theoretical side, the presence of internal rotation has strong implications on our understanding 
of the formation 
and dynamics of GCs and affects, for example, their long term evolution \citep{einsel99,ernst07,breen17} 
and their present-day morphology (e.g.,  \citealt{hong13,vdb08}). 
Moreover, signatures of internal rotation could be crucial in revealing the formation mechanisms of the so-called 
multiple stellar populations (MPs) in GCs \citep{bekki10,mastrobuono13,henault15} differing in terms of their light-element
(such as He, Na, O, C, N) abundances (see \citealt{bastian18,gratton19} for recent reviews on the subject), and which are observed in almost 
all GCs now.  
Differences in the rotation amplitudes of MPs have been observed
in two cases so far, namely M~13 and M~80 
(\citealt{cordero17,kamann20})
\footnote{Differential rotation among different populations were also observed in $\omega$ Centauri \citep{bellini18}. 
However, it is important to stress that the star formation history of this system is by far not typical for GCs and its 
sub-populations differ also in terms of age and metallicity.} and in both clusters the Na-rich population (also known as second
population or generation - SP) 
is found to rotate with a larger amplitude than the first population FP (Na-poor). 

In general GCs are characterised by moderate rotation, with typical ratios of rotational velocities to 
central velocity dispersions ($V_{rot}/\sigma_0$) ranging from about 0.05 to about 0.6 \citep{bellazzini12,fabricius14}.
However, the present-day rotation in GCs is likely 
the remnant of a stronger early rotation (see e.g. \citealt{henault12,mapelli17}) gradually weakened by 
the effects of two-body relaxation \citep{bianchini18,kamann18b,sollima19}. 
In fact, recent $N$-body simulations \citep{hong13,tiongco17} provide evidence that 
during the cluster long-term evolution, the amplitude of the rotation decreases due to 
angular momentum redistribution and star escape.
At the same time the peak of the rotation curve gradually moves toward the cluster innermost regions.

As a part of a project aimed at studying the structural and kinematical properties of 
GCs and their MPs (e.g. \citealt{dalessandro14,massari16,dalessandro18a,dalessandro18b,dalessandro19,kamann20}), our group is 
conducting a thorough investigation
of the low-mass Galactic GC NGC 6362 ($M\sim5\times10^4 M_{\odot}$). 
Based on a combination of spectro-photometric observations, our previous analyses suggest 
\citep{dalessandro14,mucciarelli16,massari17,dalessandro18a} 
that the cluster underwent severe mass-loss due to long-term dynamical evolution
and quite strong interaction with the Galaxy \citep{miholics15,kundu19}. 
In fact, we find that FP and SP stars have indistinguishable radial distributions allover the cluster extension, 
as expected for a cluster in an advanced dynamical stage \citep{vesperini13,dalessandro19}. 
In addition, the velocity dispersion profiles of the two sub-populations show differences of the order 
of $\sim1$ km/s at intermediate/large cluster-centric distances that can be ascribed to the combined effects 
of advanced dynamical evolution and a significantly larger FP binary fraction with respect 
to the SP one that can inflate the velocity dispersion by the observed amount \citep{dalessandro18a}.
The hypothesis of an advanced dynamical state of the cluster is also supported by the quite flat mass 
function of the system \citep{paust10}.  
In \citet{dalessandro18a} we also verified that NGC 6362 does not
show any significant evidence of large-scale line-of-sight 
rotation ($V_{rot}\sim0.17^{+0.32}_{-0.17}$ km/s) as also 
confirmed by \citet{bianchini18} and \citet{sollima19} by using Gaia DR2 proper motions.

Here we perform a 3D kinematic analysis of the innermost region ($R<30\arcsec$) of the cluster, that was only poorly sampled by previous 
observational campaigns. To this aim, we use a combination of proprietary MUSE 
deep data supported by adaptive optics, and HST proper motions 
published in previous works.
The paper is structured as follows. The observational data-set and data analysis 
are presented in Section~2, while in Section~3 we define the sample of stars used to perform the kinematic analysis. 
Section~4 details on the kinematic analysis and derived results and Section~5 focuses on the relative kinematics of MPs. 
Conclusions are discussed in Section~6.

\section{Observations and Data Analysis}\label{sec:obs}
The main observational data-set used in this work consists of MUSE@ESO-VLT data obtained with the support 
of the adaptive optics module GALACSI (Prop ID: 0103.D-0545; PI: Dalessandro). 
Two Wide Field Mode pointings, epoch A and B respectively, both centred on the cluster (Figure~1) and
with exposure time of 2400 sec each, were obtained between the 1st and 2nd of May 2019. 
Each pointing was observed with four different instrument derotator angles (0, 90, and 180, 270 degrees) in order to level 
out possible resolution differences between the individual spectrographs. 
The data-cubes were reduced by using the standard MUSE pipeline \citep{weilbacher20}.
An average FWHM of $\sim0.7''$ was delivered by the system for each pointing, 
whereas the uncorrected FWHM (i.e. seeing) during the observations was $\sim1.0''$.

\begin{figure}
\centering
\includegraphics[width=\columnwidth]{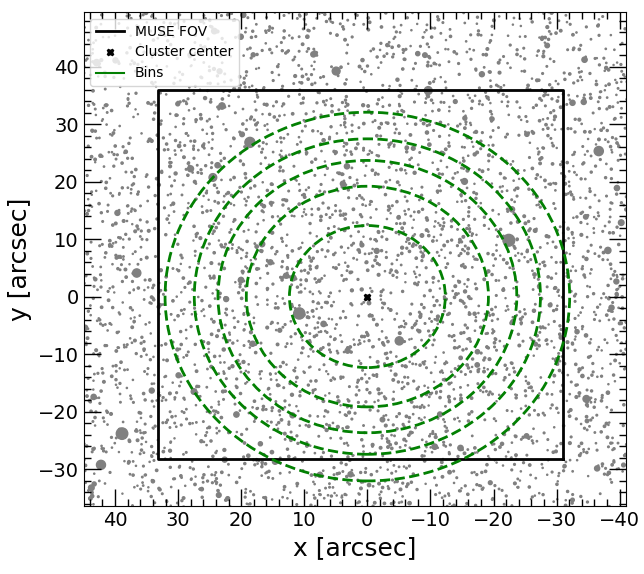}
\caption{Footprint of the MUSE observations (black box). Green circles represent the bins used for the kinematic analysis, while the black cross 
is the cluster center from \citet{dalessandro14}.}\label{fig:map}
\end{figure}

Stellar spectra were extracted from the reduced data cubes using \texttt{PampelMuse} \citep{kamann_resolving_2013,kamann_pampelmuse_2018}. 
Besides the integral-field data, the software requires a photometric reference catalogue as input. 
To this aim, we used the HST multi-band photometric catalog presented in \citet{dalessandro14}.
\texttt{PampelMuse} fits a wavelength-dependent PSF as well as a coordinate transformation from the reference HST catalogue to the MUSE data 
and uses this information to optimally extract the spectra of the resolved sources. The same analysis was performed on both the cubes created 
for the individual epochs and on the cube using all available exposures.

At the end of the analysis, we extracted $2\,658$ individual stellar spectra for epoch A and $2\,650$ individual stellar spectra for epoch B. 
While from the combined cube we were able to derive $2\,925$ spectra. 

The extracted spectra were cross-correlated against synthetic templates from the GLib library \citep{husser_new_2013} 
to derive ``first-pass'' line of sight radial velocities (LOS-RVs). 
For each extracted spectrum, a matching template was selected based on the effective temperature and surface gravity values 
derived photometrically for the corresponding star, via a comparison with an isochrone. We used a BaSTI \citep{pietrinferni06} 
$\alpha$-enhanced theoretical model with adequate metallicity ([Fe/H]=-1.09; \citealt{mucciarelli16,massari17,dalessandro18a}) 
and with age $t=12.5$ Gyr \citep{dotter10}. 
In the final step of the analysis, we performed a full-spectrum fit of each spectrum, using the \texttt{Spexxy} 
tool \citep{husser_muse_2016}. In addition to the aforementioned effective temperatures and surface gravities, 
we used in this step also the first-pass LOS-RVs as initial guesses for the fit. Given the low spectral resolution 
of the MUSE data, the surface gravities were 
held fixed at their initial guesses, while leaving the LOS-RVs, effective temperatures, and metallicities of the 
stars as free parameters.

\begin{figure}
\centering
\includegraphics[width=\columnwidth]{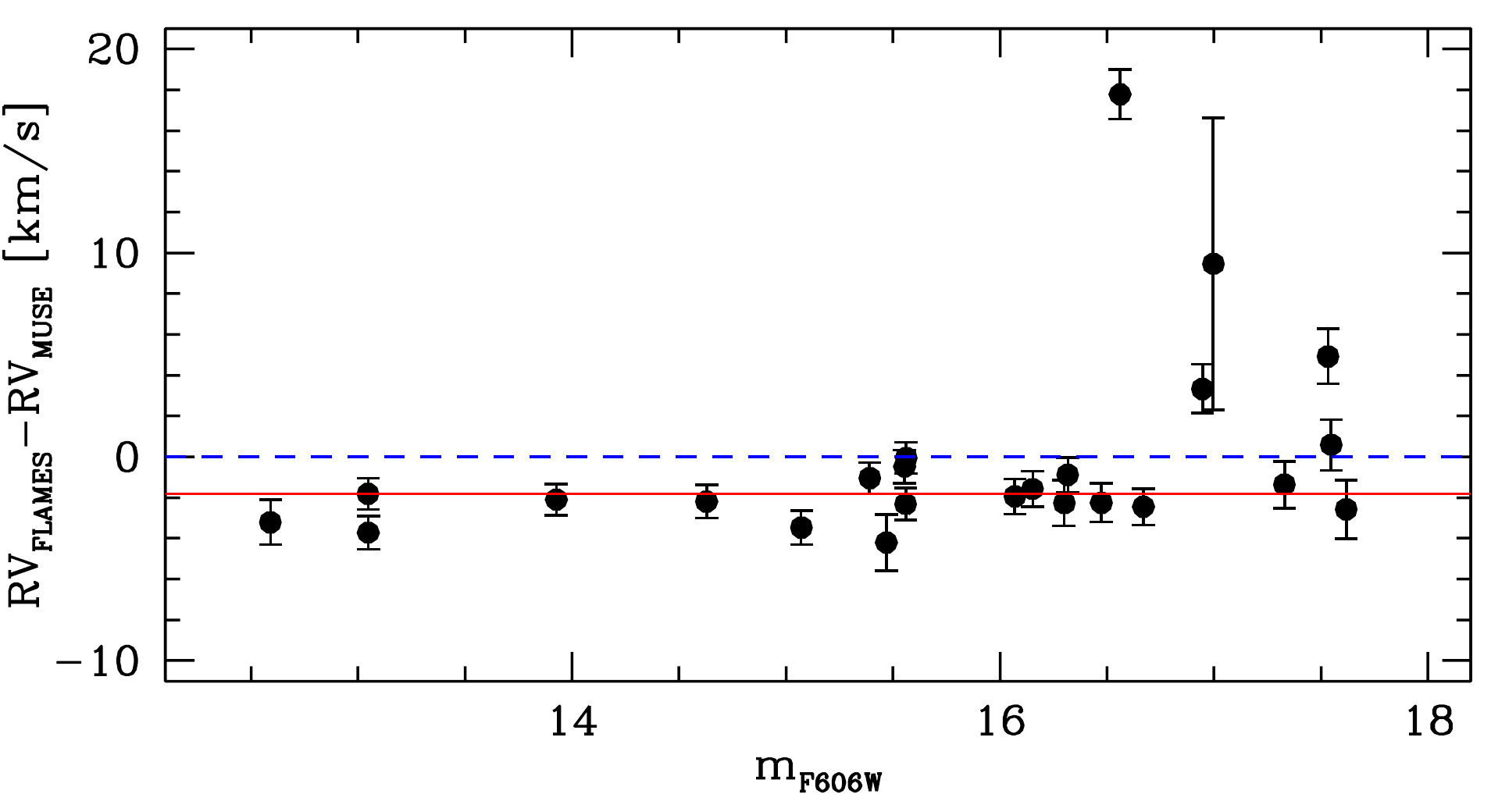}
\caption{Difference between the FLAMES and MUSE radial velocities as a function of the F606W magnitude (black circles). 
The MUSE radial velocities are obtained from the full spectral fit. The blue dashed line corresponds to 0, 
while the red solid line corresponds to the median value of the difference between the two velocities. }\label{fig:deltaflamesmuse}
\end{figure}

In addition to the MUSE catalog, we used two complementary data-sets.
We made use of the FLAMES data presented in \citet{dalessandro18a}, which cover the entire extension of the cluster out to a distance of R$\sim 800 ^{\prime \prime}$ from its center and provide high-resolution LOS-RVs for 489 cluster members selected based on both their velocities and 
metallicities (see \citealt{dalessandro18a} for details). We point out here that 
while covering a larger field of view than MUSE, the FLAMES catalog is significantly shallower and it includes only stars with $m_{F606W}<17.7$.
We checked for possible radial velocity systematic differences between the MUSE and FLAMES samples 
by using the 24 stars in common between the two datasets. In Figure~\ref{fig:deltaflamesmuse} 
we show the difference between the FLAMES and MUSE LOS-RVs as a function of the $m_{F606W}$ magnitude. 
The median value of the difference results to be $\mathrm{RV_{FLAMES}-RV_{MUSE}}$ $\sim -1.9 \ \mathrm{km \ s^{-1}}$. 
For comparison, the median value of the combined radial velocity errors for the stars in common between the two data-sets, 
is $\sim 1 \ \mathrm{km \ s^{-1}}$. 
To homogenise the two samples, and because of the larger resolution of FLAMES, we applied to all the MUSE LOS-RVs
a shift equal to the derived median difference in radial velocity. 

Finally, to study the cluster core kinematics also on the plane of the sky thus enabling a three-dimensional (3D) view, we used the 
HST proper motions (PM) catalog published by \citet{bellini14}.
We refer the reader to that papers for details about the analysis and proper motion derivation. Here we just note that the PM catalog 
extends up to distances $R\sim100\arcsec$ from the cluster center, thus totally including the MUSE field of view, and it also samples stars 
in a similar luminosity range as MUSE.

\section{Sample Selection}\label{sec:samplesel}
\begin{figure}
\centering
\includegraphics[width=\columnwidth]{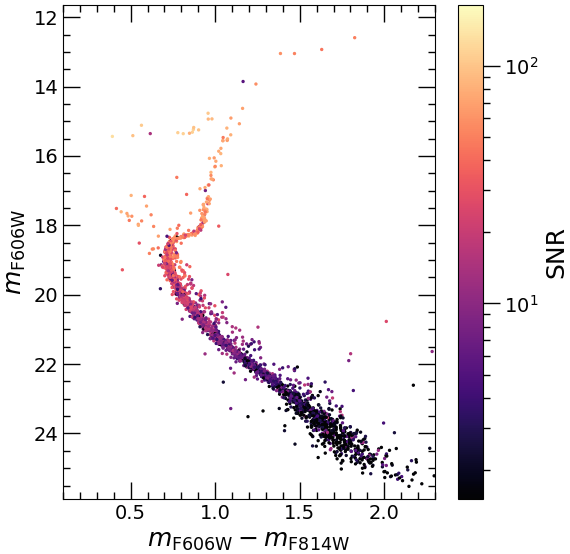}
\caption{$m_{\mathrm{F606W}}-m_{\mathrm{F814W}}$ vs. $m_{\mathrm{F606W}}$ CMD of all the stars in the MUSE catalog. 
Each star is colour-coded according to the S/N ratio of the corresponding spectrum (see colorbar on the right side 
of the Figure).}\label{fig:cmdsnr}
\end{figure}

For the kinematic analysis of the MUSE data-set, we used the LOS-RVs derived by means of the full spectrum analysis 
and obtained from the combined cube.
As shown in Figure~\ref{fig:cmdsnr}, the full sample covers a wide range of $\sim 13$ magnitudes ($12<m_{\rm F606W}<25$) 
and as consequence of signal-to-noise (S/N) ratios, which in turn range from $\sim200$ to $\sim2$. 
Since our goal is to derive kinematic quantities that are expected to have relatively small amplitudes and to compare 
the kinematic properties of different sub-populations in the cluster, we adopted rather strict selection criteria to avoid contamination 
from spurious signals of any origin.

\begin{figure}
\centering
\includegraphics[width=\columnwidth]{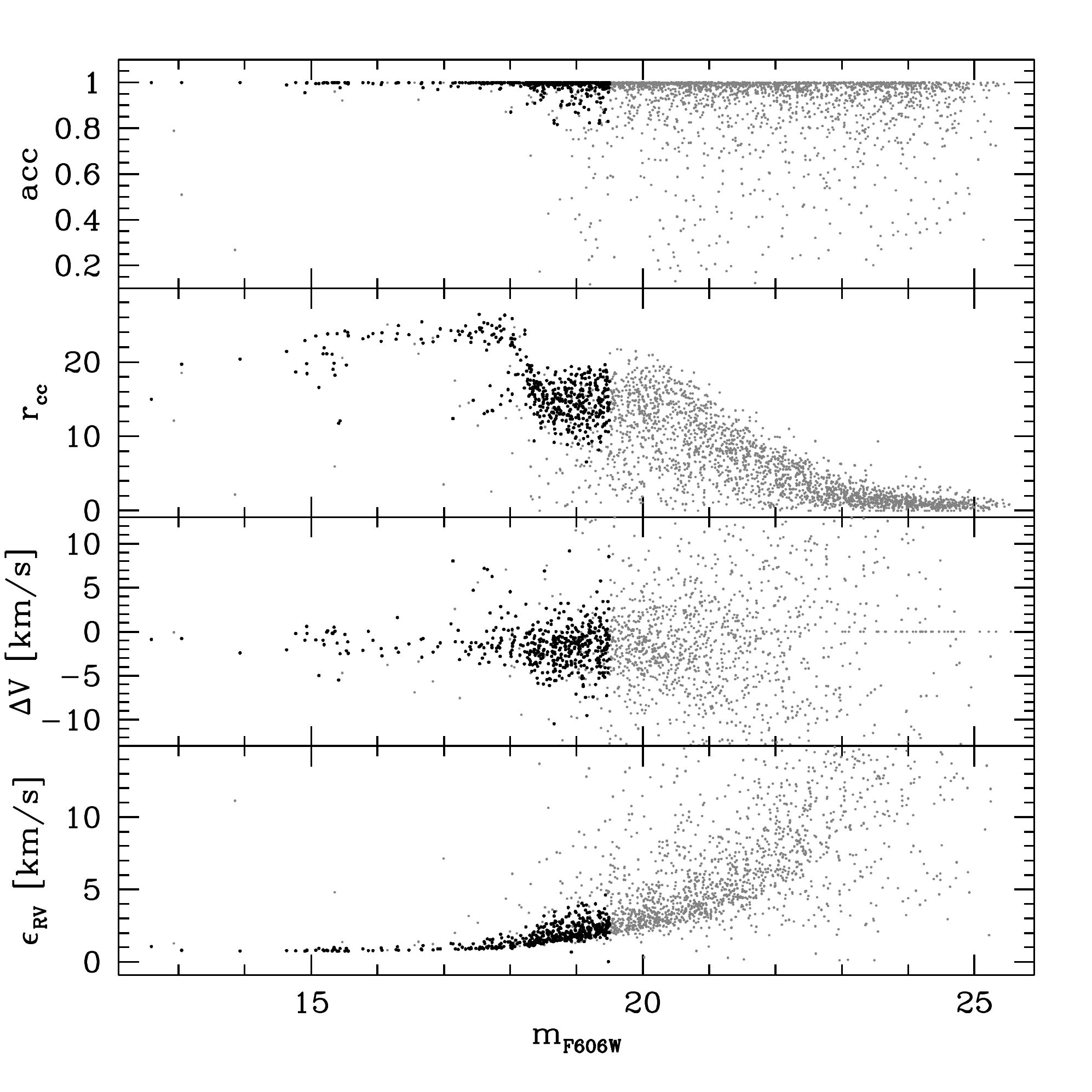}
\caption{From top to bottom: magnitude accuracy, reliability of each cross-correlation measurement 
($r_{cc}$), difference between the radial velocities measured with the cross-correlation and the full spectrum fit, 
and radial velocity error as a function of the F606W magnitude. In all panels, the complete MUSE catalog is represented 
as grey dots, while the selected sample, according to the selection criteria listed in the text, 
is represented as black dots.}\label{fig:sel_stelle}
\end{figure}

First, we excluded all stars with $S/N<15$,
which roughly corresponds to a magnitude selection $m_{\rm F606W}<19.5$.
Then, similarly to what done in \citet{kamann18a}, we adopted also the following selection criteria: {\it (i)} we defined a magnitude 
accuracy ($acc.$) as in Section 4.4 of \citet{kamann18a}, and excluded stars with $acc.\le 0.8$. This 
cut allows us to exclude stars whose LOS-RV is potentially contaminated from bright neighbours; 
{\it (ii)} we excluded stars with a $r$-parameter ($r_{cc}$ - \citealt{tonrydavis79}), which indicates the reliability of each cross-correlation measurement, 
$r_{cc}\le 4$; {\it (iii)} we measured the difference between the LOS-RVs 
obtained with the cross-correlation and with the full spectrum fit ($\Delta$v), and we rejected all the stars for which this difference 
was larger than three times the combined uncertainty of the two radial velocities measurements; 
{\it (iv)} we excluded stars that, at each given magnitude, have a LOS-RV error ($\epsilon_{RV}$) larger than $3 \sigma$ 
from the local median radial velocity error.
In Figure~\ref{fig:sel_stelle} we show the distribution of stars in the MUSE catalog as a function of the 
$m_{F606W}$ magnitude and the quality parameter just described. In all panels, we represent the complete MUSE 
sample as grey dots, and the selected sample, satisfying the criteria listed above, as black dots.

\begin{figure}
\centering
\includegraphics[width=\columnwidth]{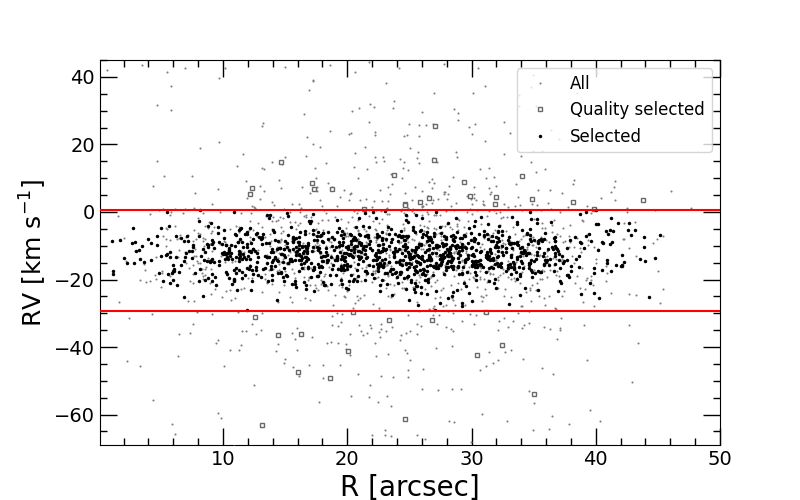}
\caption{Radial velocities as a function of the distance from the cluster center. 
The complete MUSE catalog is represented as grey dots, while the selected sample is represented as black circles. 
The two red horizontal lines mark the radial velocity cuts adopted to exclude potential non-cluster members 
(from \citealt{dalessandro18a}). The open grey squares are stars that would pass our spectroscopic quality criteria, 
but their RV is not compatible with being cluster members. }\label{fig:sel_stelle_radiusv}
\end{figure}

Finally, in order to exclude potential field interlopers in the sample of LOS-RVs selected as described above, 
following \citet{dalessandro18a}, we excluded stars with LOS-RVs outside the range: $-29.3 \ \mathrm{km \ s^{-1}} < \mathrm{RV} < 0.7 \ \mathrm{km \ s^{-1}}$. The LOS-RV distribution as function of the 
cluster-centric distance is shown in Figure~\ref{fig:sel_stelle_radiusv} along with the adopted velocity cuts to exclude likely
non-cluster members (red lines).  

The final MUSE sample surviving to the above quality and membership selections and that will be used for the following kinematic analysis counts 485 stars.
For comparison, with our previous screening with FLAMES, in the same field of view we were able to derive LOS-RVs for only 34 stars.

For the quality and membership selection of the FLAMES data we recall the reader to the detailed description in \citet{dalessandro18a}. 
Here we just stress that only FLAMES targets (465) located outside the MUSE field of view will be used in the following.
In total the final kinematic analysis will be based on a sample of reliable LOS-RVs of 950 bona-fide cluster
members.   

HST PMs were selected by using the following criteria. Only stars that {\it i)} at a given magnitude, 
have \texttt{QFIT} parameter smaller than the 80-th percentile 
(the \texttt{QFIT} parameter indicates the quality of the PSF fit, with smaller values of this parameter indicating a better quality of the 
PSF fit); {\it ii)} have PM with reduced $\chi^2,$ obtained from the PM fit, smaller than 2 and {\it iii)} 
PM measures for which the fraction of rejected displacement measurements in the PM fit procedure is smaller
than $15\%$, were considered for the following analysis. In addition, we excluded stars with a PM larger 
than 6 times the dispersion of the PM distribution to exclude obvious non-cluster members, and stars 
with a magnitude in the F606W band fainter than 19.5, 
in order to be roughly consistent with the magnitude cut indirectly applied on the MUSE data.
The final sample includes 1633 stars with reliable PMs.   

\section{Kinematic Analysis}\label{sec:kin}

The kinematic analysis was performed by using the selected sample of stars in the MUSE catalog (see Section~3) and by following 
the maximum-likelihood approach described by \citet{pryormeylan93}. The method is based on the assumption that the probability 
of finding a star with a velocity of $v_i\pm\epsilon_i$ at a projected distance from the cluster center $R_i$ can be approximated as

\begin{equation}
    p(v_i,\epsilon_i,R_i) = \frac{1}{2\pi\sqrt{\sigma^2 + \epsilon_i^2}}exp{\frac{(v_i-v_0)^2}{-2(\sigma^2 + \epsilon_i^2)}}
\end{equation}

where $v_0$ and $\sigma$ are the systemic radial velocity and the intrinsic dispersion profile of the cluster, respectively.
Rotation was included in the analysis by adding the following angular dependence (e.g., \citealt{copin01,krajnovic06,kamann18a}) 
to the mean velocity of Eq. 1.

\begin{equation}
    v_0 = v_0 + v_{rot}(R_i)sin(\theta_i - \theta_0(R_i))
\end{equation}

\begin{figure}
\centering
\includegraphics[width=\columnwidth]{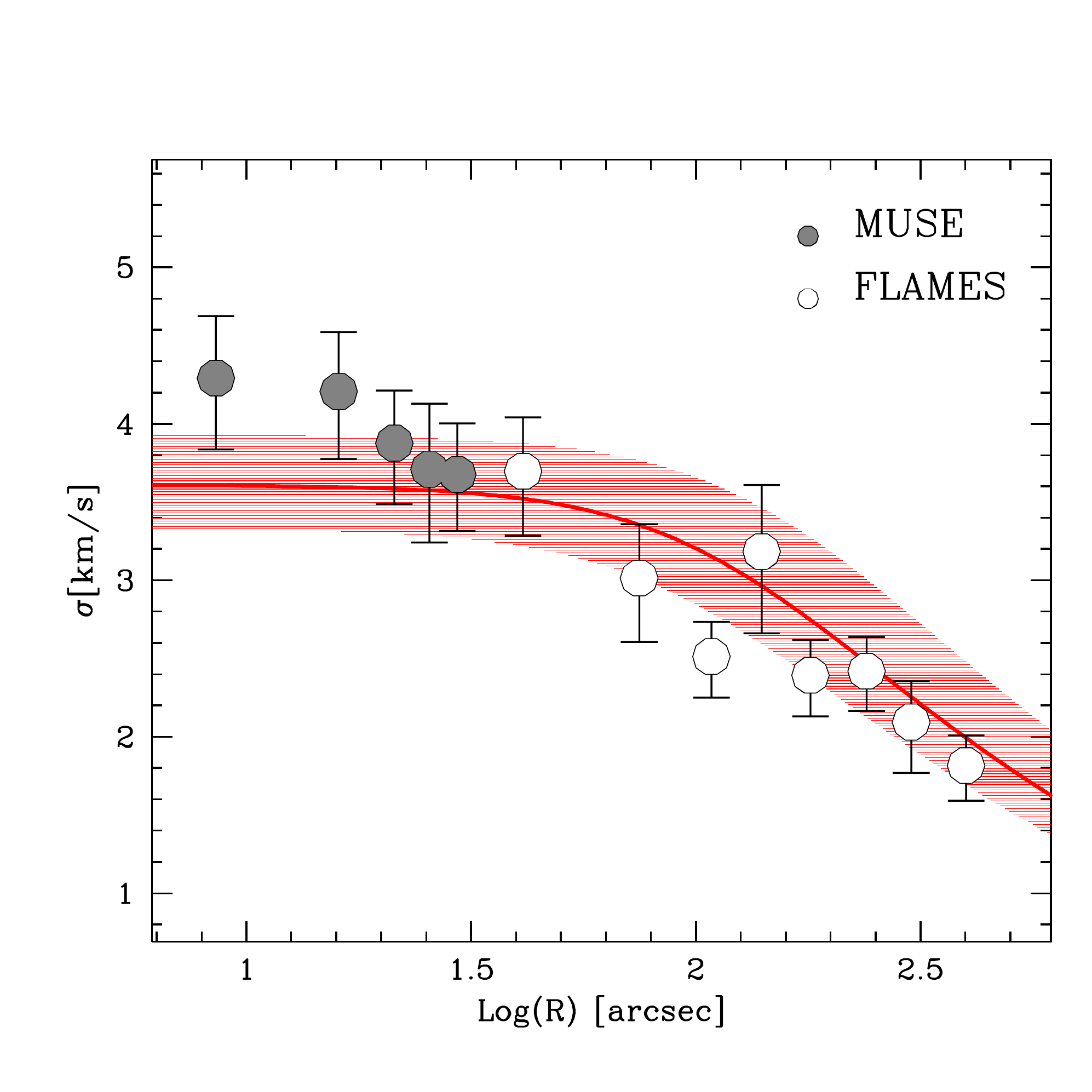}
\caption{Velocity dispersion profile obtained by the combination of MUSE (grey circles) and FLAMES (white; \citealt{dalessandro18a}) 
data.  The red line and shaded area represents the best fit curve and the corresponding error obtained in \citet{dalessandro18a}.}
\label{fig:sigTOT}
\end{figure}

where $v_{rot}$ and $\theta_0$ represent the projected rotation velocity and the rotation axis angle respectively 
as a function of the projected distance R to the cluster centre. 
The axis angle as well as the position angle $\theta_i$ of a star are measured from north through east.
We restricted the prior for $\theta_0$ 
to a $180^{\circ}$ wide interval and we allowed the rotation velocity to assume both positive and negative values, 
in order to avoid a skewed rotation velocity probability distribution in case 
of very small or no rotation (see also \citealt{kamann20}). 
In particular, we used an iterative procedure in which we adopted an angular interval 
$[\alpha; \alpha + 180^{\circ})$ centred on the most probable rotation axis angle.
 This is useful to avoid skewness in the probability distribution of $\theta_0$ when its value 
is close to $0^{\circ}$ or $180^{\circ}$.
We split the sample in five concentric annuli centred on the cluster center and with 
width varying in such a way that each bin contains the same number of stars (80).
Only stars with $r<32\arcsec$ were considered in the analysis as they guarantee an almost 
complete coverage within each annulus (Figure 1).

The fit of the kinematical quantities was performed by using \texttt{emcee} (\citealt{foremanmackey13}) with uniform priors, 
which provides the posterior probability distribution density functions (PDFs) 
for $\sigma$, $v_{\mathrm{rot}}$ and $\theta_0$ in each bin. 
We report as best-fit parameter the median of the marginalised posterior PDF of each parameter, 
and as uncertainty the interval between the 16-th and 84-th percentile of the marginalised posterior PDF.

\begin{figure*}
\centering
\includegraphics[width=\textwidth]{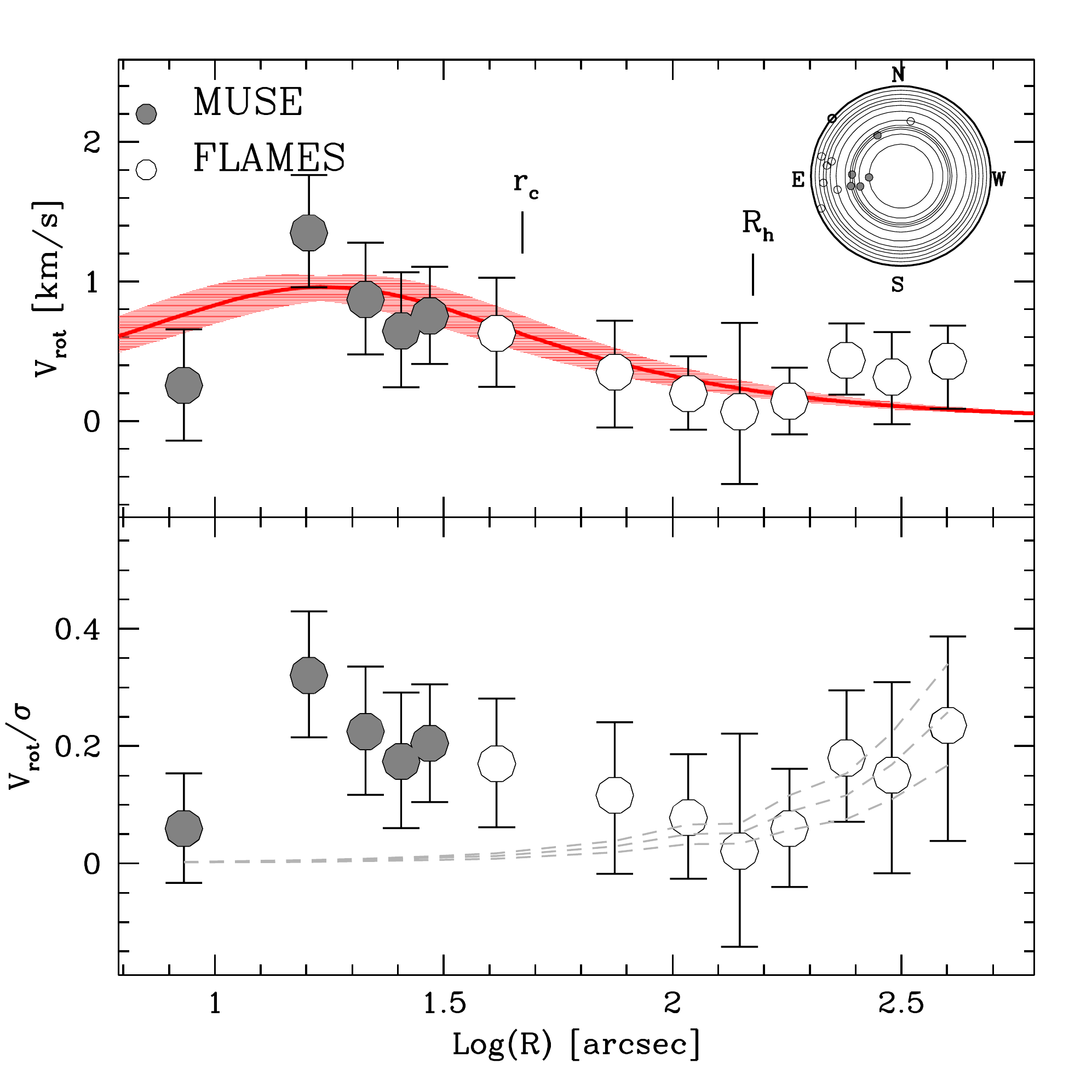}
\caption{{\bf Upper panel:} Rotational velocity and rotation axis (inset) angle as a function of the distance from the center. 
The red line and shaded area represents the best fit curve and the corresponding error, 
obtained as described in Section~\ref{sec:kin}. The core and half-light radii are also marked, for reference.
{\bf Lower panel:} $V_{rot}/\sigma$ profile. An excess reaching values of $V_{rot}/\sigma\sim0.3$ is present in the central regions also in this 
case. Grey dashed lines represent the solid-body rotation profiles obtained by assuming partial spin-orbit
synchronisation at three Galactocentric distances 
(see discussion in Section~4).}\label{fig:kinTOT}
\end{figure*}

The results of the analysis of the MUSE data are shown in 
Figures~\ref{fig:sigTOT} and \ref{fig:kinTOT} as grey circles. 
We then complemented the MUSE data with 
FLAMES LOS-RVs to study the kinematics along the entire cluster extension.
While the FLAMES sample is shallower than the MUSE one and includes only stars with $m_{F606W}<17.7$, 
it is important to note that stars in the two data-sets have approximately the same mass within $\sim0.05M_{\odot}$.
Only stars in the FLAMES catalog located at $R>35\arcsec$ were used and the same analysis as before was performed (white circles 
in Figures~\ref{fig:sigTOT} and \ref{fig:kinTOT}).

In Figure~\ref{fig:sigTOT} the resulting velocity dispersion profile is shown along with the best-fit Plummer model 
obtained in \citet{dalessandro18a}.
We note that the MUSE data show some hints of an increasing dispersion in the innermost two bins, however they are still compatible with 
the expected flat behaviour.

\begin{table}
\begin{footnotesize}
\begin{center}

\setlength{\tabcolsep}{0.2cm}

\caption{MUSE + FLAMES velocity dispersion and rotation profiles.}
\label{PROF}
\begin{tabular}{rccc}
\hline
\hline
   R 	         & $\sigma$        &   $V_{rot}$      &         $\theta_0$   \\      
                 &  [km/s]          &       [km/s]     &      [degrees] \\
\hline
 $8.5''$  &   4.29$^{+0.45}_{-0.40}$ & 0.25$^{+0.40}_{-0.40}$ &  92$^{+60}_{-56}$    \\
$16.0''$  &   4.21$^{+0.43}_{-0.38}$ & 1.35$^{+0.39}_{-0.41}$ & 104$^{+24}_{-24}$  \\
$21.3''$  &   3.88$^{+0.39}_{-0.34}$ & 0.87$^{+0.39}_{-0.41}$ &  30$^{+44}_{-37}$     \\
$25.5''$  &   3.71$^{+0.47}_{-0.42}$ & 0.65$^{+0.40}_{-0.42}$ &   88$^{+62}_{-59}$    \\
$29.5''$  &   3.68$^{+0.36}_{-0.33}$ & 0.75$^{+0.34}_{-0.35}$ & 101$^{+40}_{-42}$   \\
$41.3''$  &   3.69$^{+0.41}_{-0.34}$ & 0.63$^{+0.38}_{-0.40}$ & -10$^{+47}_{-58}$     \\
$74.8''$  &   3.01$^{+0.41}_{-0.34}$ & 0.35$^{+0.40}_{-0.37}$ & 102$^{+65}_{-47}$       \\
$108.3''$ &   2.51$^{+0.26}_{-0.22}$ & 0.20$^{+0.26}_{-0.27}$ &  78$^{+52}_{-70}$  \\
$140.1''$ &   3.18$^{+0.52}_{-0.43}$ & 0.07$^{+0.51}_{-0.63}$ &  82$^{+62}_{-76}$ \\
$180.0''$ &   2.39$^{+0.26}_{-0.23}$ & 0.14$^{+0.24}_{-0.24}$ &  95$^{+59}_{-52}$ \\
$239.9''$ &   2.42$^{+0.25}_{-0.22}$ & 0.43$^{+0.25}_{-0.27}$ &  76$^{+45}_{-57}$ \\
$301.5''$ &   2.09$^{+0.32}_{-0.26}$ & 0.31$^{+0.34}_{-0.32}$ & 112$^{+82}_{-46}$ \\
$399.5''$ &   1.81$^{+0.23}_{-0.19}$ & 0.43$^{+0.33}_{-0.25}$ &  50$^{+31}_{-96}$  \\
\hline
\hline

\end{tabular}

\small
{\bf Note:} For each annulus the table lists the the mean radius (R), the velocity dispersion ($\sigma$),
the rotation velocity ($V_{rot}$), the position angle ($\theta_0$) and relative errors.

\end{center}
\end{footnotesize}
\end{table}

As for the rotation (Figure~\ref{fig:kinTOT}), a significant signal is clearly visible when considering the MUSE data.
In fact, the profile increases out to Log(R)$\sim1.2$ ($\sim20\arcsec$) 
from the cluster center showing a maximum amplitude of $\sim1.4$ km/s. 
Then it tends to slowly decline moving outward.  
The addition of the FLAMES data confirms the smoothly declining
external branch of the rotation profile. 
The mean rotation axis angle calculated by averaging all the single bin values results to be located approximately at 
$\theta_0\sim77^{\circ}\pm35^{\circ}$, where the error is given by the standard deviation. 
The derived MUSE + FLAMES velocity dispersion and rotation values are reported in Table~1.

In Figure~\ref{fig:kinTOT} we also show the best fit curve of the rotational velocity radial profile, 
obtained imposing that it has the following analytic form, as appropriate for cylindrical rotation \citep{lynden-bell67}:

\begin{equation}
v_{\mathrm{rot}} = \frac{2 {v_{pk}} R}{{R_{pk}}} \Bigl(1+\Bigl(\frac{R}{{R_{pk}}}\Bigr)^2\Bigr)^{-1}
\end{equation}

where $R_{pk}$ and $v_{pk}$ represent the location of the rotation peak and its amplitude respectively.
The best fit curve was obtained by letting the values of $v_{pk}$ and $R_{pk}$ vary and
by estimating the reduced $\chi^2$ of the residuals between the observed and the model profiles. 
The solution corresponding to the lowest value of the stored $\chi^2$ is finally adopted as the best-fit model. 
The errors on these two parameters correspond to the interval where $\chi^2 \le \chi^2_{\mathrm{best}}+1$. 
The derived best fit values are $v_{pk} = (0.96 \pm 0.09)$ km/s 
and $\mathrm{R_{pk}} = (17.3^{+2.7}_{-2.8}) ^ {\prime \prime}$.

In the lower panel of Figure~\ref{fig:kinTOT} we also show the $V_{rot}/\sigma$ profile that provides a direct measure 
of the ordered to random 
stellar motion. A peak at $V_{rot}/\sigma=0.32\pm0.1$ is visible in this plot at the same bin as the rotation profile peak. 

{\it This result represents the first detection of rotation in NGC~6362. In addition, 
with a peak located at only $\sim 0.1 R_h$ ($R_h=150\arcsec$; \citealt{dalessandro14}) 
this is one of the most central rotation signal ever observed in a GC to date.} A similar case was 
found in the massive post-core-collapse GC M~15, which shows a decoupled rotating core \citep{vdb06,usher21}.
Likely the combination of a modest absolute amplitude and the very central position of the rotation peak has made the signal
elusive to previous screenings.

We also note that $V_{rot}/\sigma$ tends to increase in the cluster outermost regions ($Log(R)>2.2$). 
$N$-body simulations of the long-term evolution of GCs have shown that clusters evolve toward a state of partial 
spin-orbit synchronisation characterised by an internal solid-body rotation with angular velocity equal to $\Omega/2$ where $\Omega$ 
is the angular velocity of the cluster's orbital motion around the Galactic center \citep{tiongco16}. 
In the lower panel of Figure 7 we have plotted three radial profiles for $V_{rot}/\sigma$ with $V_{rot}$ 
calculated assuming a solid-body rotation with angular velocities $\Omega_p/2$, $\Omega_a/2$, and $\Omega_e/2$ 
calculated by simply assuming $V_{circ}=220$ km/s and Galactocentric distances equal to, respectively, the cluster's pericenter 
($R_p=2.54$ kpc), apocenter ($R_a=5.16$ kpc) and effective Galactocentric distance 
($R_e=R_a(1-e)$ where $e=0.35$ is the orbital eccentricity; see \citealt{baumgardt19}). 
The solid-body rotation profiles shown in Figure~\ref{fig:kinTOT} suggest that the outer kinematic properties revealed by our data  
might represent the signature of the cluster's partial spin-orbit synchronisation. 
Additional data allowing a firmer determination of the cluster's 
outer kinematics along with specific models aimed at modelling the kinematic evolution of NGC 6362 are however necessary 
to further explore this issue.

\subsection{Proper motion analysis}
We also studied the kinematics of the central regions of NGC~6362 in the plane of the sky by using the PM catalog described in Sections~2 and 3 for
stars with $m_{F606W}<19.5$. 
We stress here that because of the way proper motions are derived, it is not possible to quantitatively study the rotation 
as it is (at least partially) erased by the linear transformations necessary to report all catalogs on the same astrometric 
reference frame. However, some residuals can be expected to be still imprinted along the tangential component of the motion.    

We divided the selected PM sample into six equally populated (270 stars per bin) radial bins. 
In each bin we derived the velocity dispersions in both the tangential and radial components by using again a maximum likelihood approach 
and the \texttt{emcee} (\citealt{foremanmackey13}) algorithm to obtain the posterior PDFs for the tangential and radial components
of the velocity dispersion $\sigma_{tan}$ and $\sigma_{rad}$ respectively (see \citealt{raso20} for more details).

\begin{figure}
\centering
\includegraphics[width=\columnwidth]{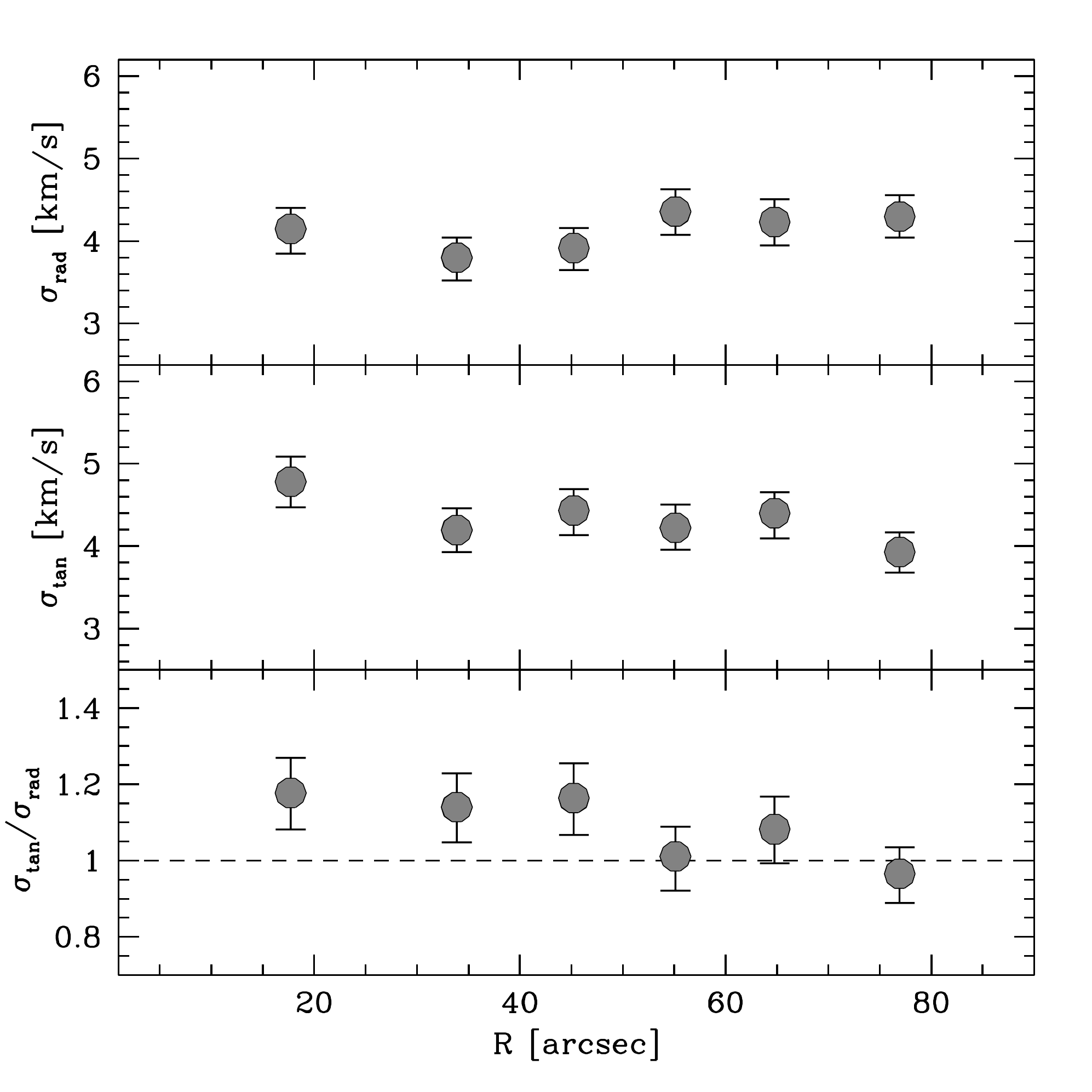}
\caption{Kinematic analysis in the plane of the sky as obtained 
by using the quality selected PM sample and dividing it into six equally populated bins. 
From top to bottom: radial and tangential components of the velocity dispersion in the plane of the sky 
and anisotropy profile, obtained as the ratio between the tangential and radial components. 
The dashed line in the bottom panel marks isotropy. }\label{fig:kinTOT_PM}
\end{figure}

The results of this analysis are summarised in Figure~\ref{fig:kinTOT_PM}.
The anisotropy profile (defined here as $\sigma_{tan}/\sigma_{rad}$) shows clear evidence of an excess
of motion along the tangential component in the innermost $30\arcsec-40\arcsec$.
Indeed, the innermost three radial bins have an average anisotropy value $<\sigma_{tan}/\sigma_{rad}>=1.16\pm0.02$ 
and each of the three measures exceeds the isotropic behaviour by $\sim2\sigma$.   
We note that a hint of central tangential anisotropy was also found by \citet{watkins15}.
This behaviour is in qualitative agreement with what 
expected based on the central rotation observed 
with MUSE LOS-RVs. 

To perform a one-to-one comparison with the MUSE results, 
we repeated the analysis including only stars (453) in common with the selected MUSE catalog. 
Also in this case, 
the PM analysis reveals a consistent indication of tangential anisotropy in the innermost regions of the cluster although the 
distribution results to be noisier because of the lower number statistics.

\subsection{Ellipticity}
\begin{figure}
\centering
\includegraphics[width=\columnwidth]{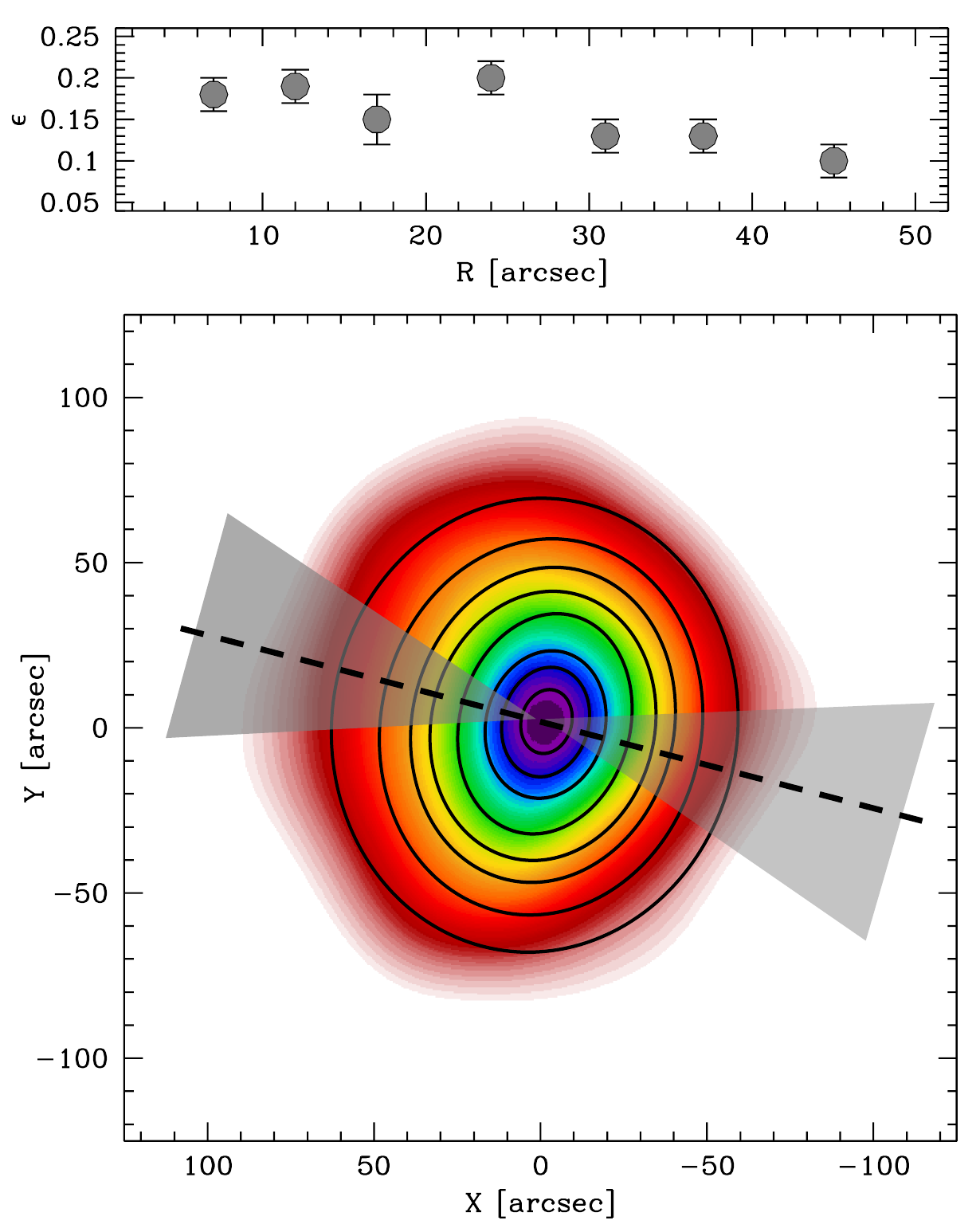}
\caption{{\bf Lower panel:} smoothed stellar density map of the inner $85\arcsec\times85\arcsec$ of NGC 6362, obtained from HST photometry. 
The solid black lines represent the best-fit ellipses to the isodensity curves, the dashed lines represents
the average rotation axis obtained from the kinematic analysis and the shaded area the relative uncertainties. {\bf Upper panel:} ellipticity distribution as a function of the cluster-centric distance.}
\label{fig:dens_map}
\end{figure}

A rotating system is also expected to be flattened in the direction perpendicular to the rotation axis \citep{chandrasekhar69}.
We used the HST catalog from \citet{dalessandro14} to construct a two-dimensional (2D) density map of the cluster central regions and study its morphology. 
The 2D density map shown in Figure~\ref{fig:dens_map} (lower panel) was obtained by transforming the distribution of stars $m_{F606W}<25$ 
into a smoothed surface density function through the use of a gaussian kernel with width of $5\arcsec$. 
To minimise the effect of the limited field of view on the smoothing procedure, the analysis was limited 
to an area of $\sim 85\arcsec\times85\arcsec$.
The lower panel of Figure~\ref{fig:dens_map} also shows the best-fit ellipses to the isodensity contours. 
The distribution of their ellipticity $\epsilon=1 - b/a$ where $a$ and $b$ are the major and minor axis respectively, 
as a function of the cluster-centric distance are shown in the upper panel and reported in Table~2.   
As apparent, the ellipticity is more prominent in the innermost rotating regions and it reaches its maximum 
($\epsilon=0.20\pm0.02$) at $R\sim20\arcsec$. Then ellipticity progressively smooths out moving outward. 
In fact, in the outermost regions $\epsilon=0.1\pm0.02$, consistently with what found at larger distances 
(see for example \citealt{harris96}). The ellipses major axis tend to have an orientation of $\sim160^\circ$ in the North-East
direction and the stellar density distribution is flattened in the direction of the average 
position angle of the rotation axis consistent, in general, with the expectation for a system 
flattened by its internal rotational velocity. 

\begin{table}
\begin{footnotesize}
\begin{center}

\setlength{\tabcolsep}{0.2cm}

\caption{Ellipticity profile}
\label{ELL}
\begin{tabular}{rccc}
\hline
\hline
   R 	         & $\epsilon$   &   $\sigma_{\epsilon}$      &   orientation  \\      
       &                     &          &                         [degrees]    \\
\hline
 7$''$  &   0.18 & 0.02 &  155$\pm$15   \\
12$''$  &   0.19 & 0.02 &  157$\pm$13  \\
17$''$  &   0.15 & 0.03 &  165$\pm$13     \\
24$''$  &   0.20 & 0.02 &  165$\pm$12     \\
31$''$  &   0.13 & 0.02 &  162$\pm$14  \\
37$''$  &   0.13 & 0.02 &  160$\pm$15    \\
45$''$  &   0.10 & 0.02 &  160$\pm$15      \\

\hline
\hline

\end{tabular}

\small
{\bf Note:} For each annulus the table lists the mean radius (R), the ellipticity value ($\epsilon$),
the orientation of the ellipse major axis and relative errors.

\end{center}
\end{footnotesize}
\end{table}

\section{Kinematics of Multiple Populations}\label{ss:sepmp}
MPs are believed to form during the very early epochs of GC formation and evolution ($10 - 100$ Myr). 
A number of scenarios have been proposed over the years to explain their formation, however their origin 
is still strongly debated \citep{decressin07,dercole08,bastian13,denissenkov14,gieles18,calura19}.
It has been shown that the kinematical and structural properties of MPs can provide key insights into the early epochs of GC evolution and formation. 
In fact, one of the predictions of MP formation models (see e.g. \citealt{dercole08}) is that SP stars form a centrally 
segregated stellar sub-system possibly characterised by a more rapid internal rotation \citep{bekki11} 
than the more spatially extended FP system. 
Although the original structural and kinematical differences between FP and SP stars are gradually erased during GC long-term 
dynamical evolution (see e.g. \citealt{vesperini13,henault15,tiongco19}), some clusters are expected to still retain 
some memory of these initial differences in their present-day properties (e.g. \citealt{richer13,bellini15,
dalessandro16,cordero17,dalessandro18b,dalessandro19,kamann20}).
Therefore connecting the kinematic and chemical properties of multiple populations in GCs may offer a 
valid approach for understanding how they 
formed.

\begin{figure}
\centering
\includegraphics[width=\columnwidth]{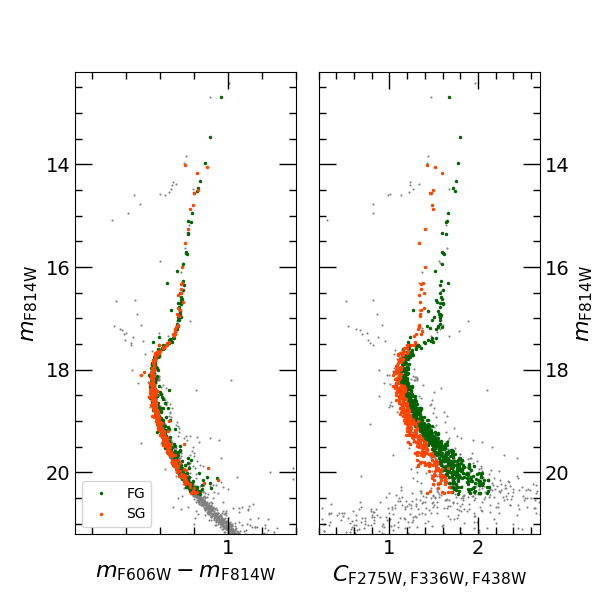}
\caption{{\bf Left panel:} ($m_{\mathrm{F814W}}$, $m_{\mathrm{F606W}}-m_{\mathrm{F814W}}$) CMD. 
{\bf Right panel:} ($m_{\mathrm{F814W}}$, $C_{\mathrm{F275W,F336W,F438W}}=(m_{\mathrm{F275W}}-m_{\mathrm{F336W}})-(m_{\mathrm{F336W}}-m_{\mathrm{F438W}})$) 
pseudo-CMD. In both panels, all the stars from the complete MUSE catalog are shown as grey dots, while (quality selected, 
see Section~\ref{sec:samplesel}) FP and SP stars are shown as green and orange dots, respectively.}\label{fig:cmd_selMP}
\end{figure}

The kinematic properties of NGC 6362 MPs have been investigated in detail in \citet{dalessandro18a}. 
At odds with the expectations for a dynamically evolved cluster 
whose MPs share the same radial distributions \citep{dalessandro14}, we found that at distances from 
the cluster center larger than about 0.5$R_h$, FP and SP stars show hints of different 
line-of-sight velocity dispersion profiles, with FP stars being dynamically hotter. 
This effect is likely due to 
a significant difference of the relative binary fraction of FP ($\sim15\%$) and SP ($\sim1\%$) stars. 
On the contrary, we did not find any evidence of different rotation between the two populations.  
Here we take advantage of the exquisite MUSE performance and the large sample of stars with available LOS-RVs 
in the innermost $30\arcsec$ to investigate this aspect further.

\begin{figure*}
\centering
\includegraphics[width=\textwidth]{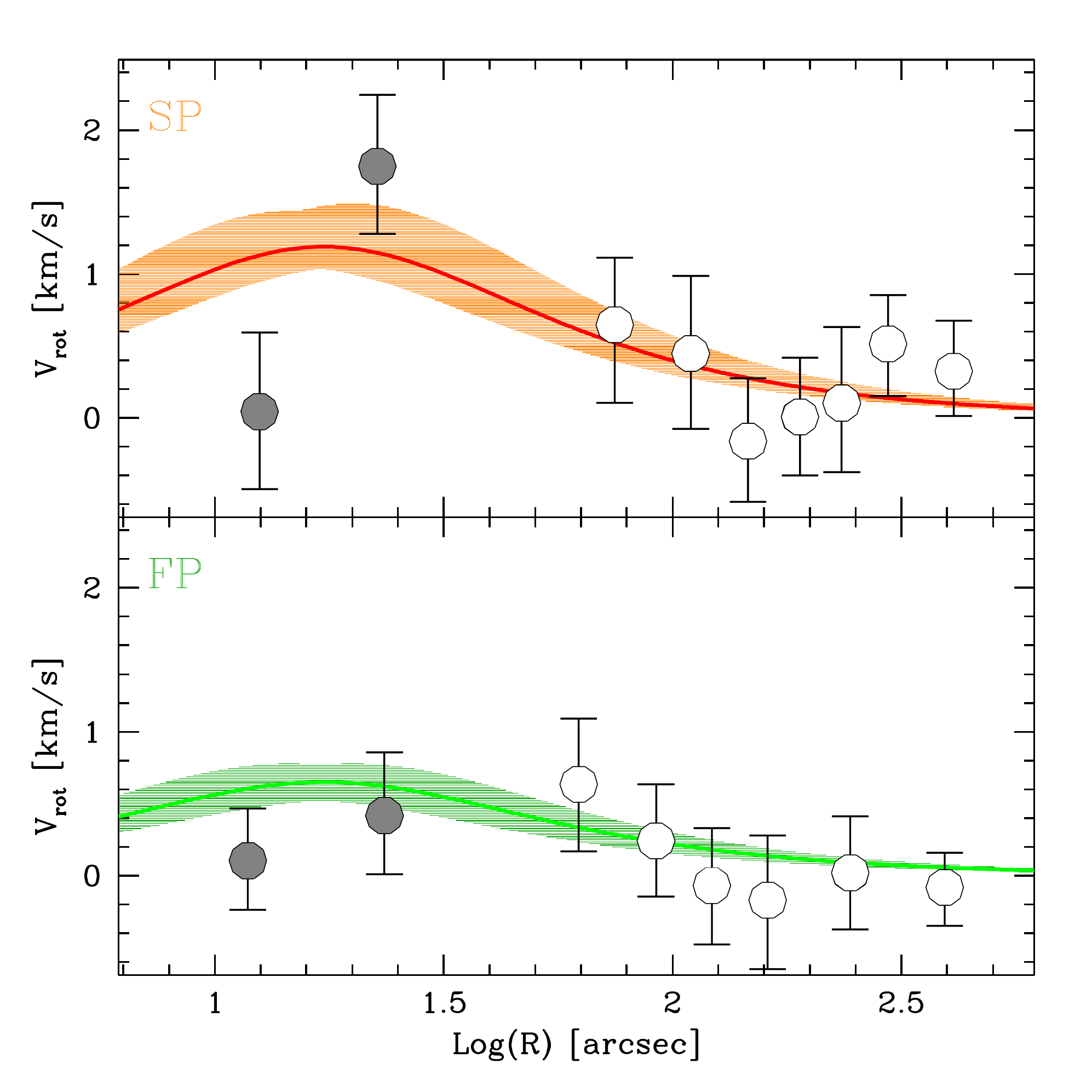}
\caption{Rotational velocity profiles from MUSE and FLAMES data for FP and SP stars (top and bottom panels, respectively). 
We also show the best fit curves (see text for details about the fit) and relative errors.}\label{fig:kinMP}
\end{figure*}

In order to separately analyse the two populations of NGC~6362, corresponding to different chemical abundances, 
we cross-correlated the positions and magnitudes of stars in our MUSE catalog with the ones from 
the \textit{HST} UV Legacy Survey of 
Galactic GCs\footnote{Catalogs available at \url{http://groups.dfa.unipd.it/ESPG/treasury.php}.} 
(\citealt{piotto15,nardiello18}). UV bands are
efficient in separating MPs in CMDs, as variations of the OH, CN, and CH molecular bands have strong
effects in the spectral range $3000 - 4000 \AA$ \citep{sbordone11}.
Using the F275W, F336W, F438W magnitudes, combined with the F814W band, we constructed adequate color-color diagrams 
to distinguish between FP and SP stars along the red giant branch (RGB), sub-giant branch (SGB), 
and main sequence (MS; \citealt{milone20}). 
In particular, we divided our sample into three $m_{\mathrm{F814W}}$ magnitude bins, 
separated at $m_{\mathrm{F814W}}=18.1$ and $m_{\mathrm{F814W}}=19.3$ thus separating the RGB, SGB and MS. 
For the RGB and MS, we constructed the so-called ``chromosome map'' \citep{milone17} 
defined as the 
$(m_{\mathrm{F275W}}-m_{\mathrm{F814W}}$,
$C_{\mathrm{F275W,F336W,F438W}}=(m_{\mathrm{F275W}}-m_{\mathrm{F336W}})-(m_{\mathrm{F336W}}-m_{\mathrm{F438W}})$) 
color-color diagram.  Briefly, 
we verticalized the star distribution in the $m_{\mathrm{F814W}}$ vs. $C_{\mathrm{F275W,F336W,F438W}}$ pseudo-CMD and in the $m_{\mathrm{F814W}}$ vs. $m_{\mathrm{F275W}}-m_{\mathrm{F814W}}$ CMD with respect to two fiducial lines at the blue and red edges of the sequence. The chromosome map 
corresponds to the combination of the two verticalized distributions. For the SGB, we instead constructed the $m_{\mathrm{F336W}}-m_{\mathrm{F438W}}$ vs. $m_{\mathrm{F275W}}-m_{\mathrm{F336W}}$ color-color diagram. In the appropriate diagram for each bin, the star distribution appears to be bimodal and, therefore, the two populations can be easily separated.

In Figure~\ref{fig:cmd_selMP} we show the ($m_{\mathrm{F814W}}$, $m_{\mathrm{F606W}}-m_{\mathrm{F814W}}$) CMD 
(left panel) and the ($m_{\mathrm{F814W}}$, $C_{\mathrm{F275W,F336W,F438W}}$) pseudo-CMD (right panel), 
where we highlight FP stars in green and SP stars in orange.

For the FLAMES data-set FP and SP stars were separated by using the Na abundances obtained in \citet{mucciarelli16} and 
\citet{dalessandro18a} and using a separation limit at [Na/Fe]=0.05 (stars with [Na/Fe]$<0.05$ are classified as FP, while 
Na-rich stars ar SP). The nice match between UV color separation 
and [Na/Fe] abundances was also shown in detail in those papers.

We then repeated the kinematic analysis described in Section~4 for the FP and SP sub-populations separately.
We reduced the total number of equally populated bins, given the smaller total number of stars in each stellar sub-group. 
As we did previously, we discarded the outermost bin because of the lack of circular symmetry due to the shape of the MUSE FOV, 
that could potentially bias the analysis. The results of the MP kinematic analysis are summarised 
in Figure~\ref{fig:kinMP}. In the top and bottom panels we show the rotational velocity profiles obtained for SP and FP stars
respectively. 
As in Figure~\ref{fig:kinTOT}, grey circles correspond to the MUSE data and open circles to the FLAMES data. 
We also show the best fit curve, obtained in the same way as for the total population. 
We should stress here, that because of the lower number of radial bins and larger uncertainties, 
we decided to adopt the same value of $R_{pk}$ as obtained for the entire sample in the fitting procedure.
The SP population shows the stronger rotation with a peak velocity $v_{pk}=1.19^{+0.25}_{-0.14}$, while the FP rotates at a 
lower pace with $v_{pk}=0.65^{+0.13}_{-0.12}$. The difference in rotation velocity between MPs is significant 
at a $2.8\sigma$ level.  
The different binary fraction between FP and SP stars, which has been shown 
\citep{dalessandro18a} to have an impact 
on the cluster kinematics at intermediate-large cluster-centric distances, is not expected to play a role
on the differential rotation detected here (see also \citealt{hong19}).
Also, we note that the slightly increasing profile (at $Log(R)>2$) of $V_{rot}$ and $V_{rot}/\sigma$ discussed above seems to be driven mainly by the SP population. 

Additional data are necessary to draw firmer conclusions concerning this feature and the possible differences in its significance 
in the SP and the FP populations. Further insight on these kinematic properties can provide key constraints on the dynamical 
evolution of multiple populations. 
Finally we find that the average rotation axis angle of the two populations are $102^{\circ} \pm40^{\circ}$ and $120^{\circ} \pm36^{\circ}$ 
for the FP and SP respectively, thus consistent with that of the total sample. 

\begin{figure}
\centering
\includegraphics[width=\columnwidth]{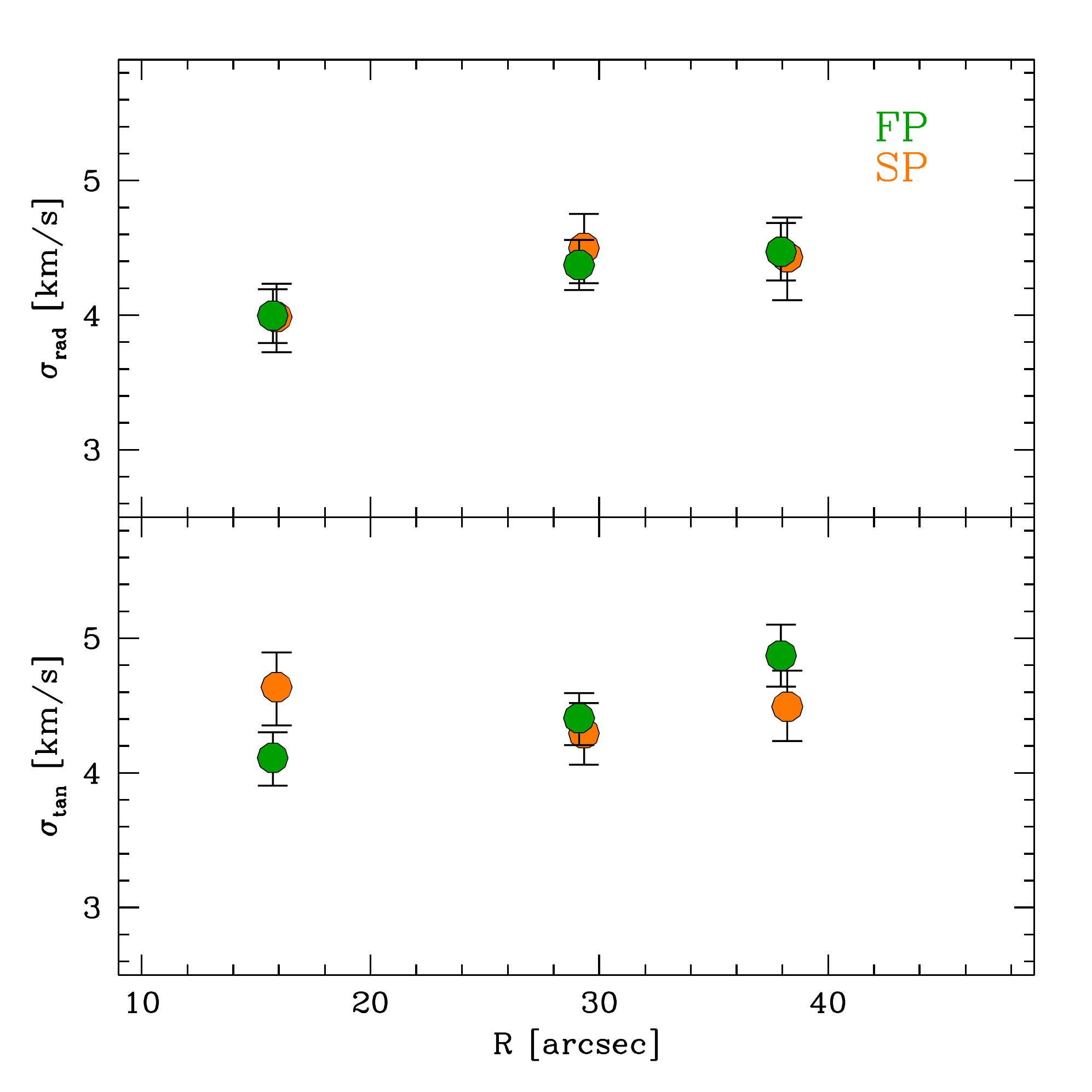}
\caption{From top to bottom: radial and tangential velocity dispersion profiles in the plane of the sky, 
obtained from PM, for FP and SP stars (green and orange symbols, respectively). }\label{fig:kinMP_PM}
\end{figure}

We matched the \textit{HST} PM catalog with the FP and SP stars to study their kinematic patterns also in the plane of the sky. 
We divided the two populations in three equally populated radial bins and we studied the radial and 
tangential components of the velocity dispersions. 
Results are shown in Figure~\ref{fig:kinMP_PM}.
The two populations show a very similar behaviour in terms of their radial components as expected for two sub-populations sharing the
same radial distributions \citep{dalessandro14}. On the contrary, the SP shows a larger tangential dispersion 
than FP stars in the innermost bin that can be interpreted as evidence on the plane of the sky of a larger rotation, 
in agreement with the difference 
in the rotation profile found with the MUSE data. 

The observed MP 3D kinematic results are generally consistent with a formation scenario in which 
a second population formed from the ejecta of a rotating first population (see e.g. \citealt{bekki10,bekki11}). 
These results are also in qualitative agreement with the studies of M 13 by \citet{cordero17} and M 80 by \citet{kamann20}, 
who also found a higher rotation velocity for the SP sub-population.

\section{Summary and discussion}
The high-resolution 3D kinematic analysis presented in this paper has revealed for the first time a significant rotation signal 
in the innermost regions ($R<30\arcsec$) of NGC~6362. 
The rotation curve shows a velocity peak $v_{pk}\sim1$ km/s at $R_{pk}\sim 20\arcsec$ roughly corresponding 
to $0.1 R_h$, then the rotation signal rapidly declines and totally disappears for $R>100\arcsec$.
Such a central rotation peak is a rare feature in GCs, with the only comparable case being M15 
(e.g. \citealt{vdb06}). However, it is worth stressing that 
at odds with NGC 6362, M 15 is a post-core collapse cluster and the observed peculiar rotation patterns can be due 
to recent physical events like cluster oscillations (see for example \citet{usher21}.

Based on results from $N$-body simulations \citep{tiongco17}, a very central rotation peak ($R_{pk}<R_h$) 
is expected only in very dynamically evolved star clusters that lost a significant amount of their mass because 
of both two-body relaxation effects and interaction with the host galaxy potential. 
In fact, in rotating stellar clusters the peak of the rotation curve is expected to be located between 1 and 2 $R_h$  
during the first stages of the cluster's evolution and then to gradually move inward in the very advanced dynamical phases.

Rotation is also expected to become weaker as the system evolves and loses mass and the current observed rotation 
is therefore only the remnant of a stronger primordial one. Indeed, assuming that 
NGC 6362 lost $50\%-80\%$ of its initial mass as constrained by its MP radial distributions \citep{dalessandro14}, 
the simulations presented in \citet{tiongco17} would suggest the initial rotation could have been $5-10$ 
times larger than the current one.

While the observed values of both $R_{pk}$ and $v_{pk}$ are in general qualitative agreement with the expected dynamical 
evolution for a star cluster, the value of the peak of $V_{rot}$/$\sigma$ we find ($\sim0.3$; see Figure 7) 
is larger by a factor of about 5 than the value suggested by simulations for clusters in the advanced stages 
of their evolution like NGC 6362. 

The derived values $v_{pk}$/$\sigma$ and the location of $R_{pk}$ possibly suggest that NGC~6362 formed with
initial conditions characterised by a high rotation and/or that internal dynamical processes may have been able to
preserve the cluster central rotation for a long timescale.
Additional simulations will be needed to understand the range of initial kinematic properties and the dynamical
ingredients required to explain such a high value of the peak of $V_{rot}$/$\sigma$ in the late stages of a
cluster's evolution.

We can in principle exclude that the very different binary fraction 
between FP and SP sub-populations is playing a role, as they are expected to manifest their effects at intermediate-large 
distances from the cluster center \citep{dalessandro18a,hong19}. 
It might be interesting to consider that 
MPs, which were likely born with different primordial structural 
and kinematic patterns, contribute in different ways to the overall energy budget of the cluster. 
In this context, it is worth emphasising that in NGC~6362 the SP sub-population is actually dominating the total rotation 
signal in the innermost region and recent observations \citep{cordero17,kamann20} 
suggest this might actually be a common behaviour. 

Further investigations both from the observational and theoretical point of view are certainly needed to shed light on both
the formation of GCs and their MPs and how the co-existence of sub-populations of stars 
with different initial kinematic properties may impact our understanding 
of the kinematics and structure of present-day star clusters. The simplicity of GCs is certainly a concept of the past,
and while unprecedented observations unveils more and more details,
theoretical models accounting for both kinematical and chemical complexities of GC stellar populations become more and more
important.

\section*{Acknowledgements}
The authors thank the anonymous referee for the careful reading of the paper 
and the useful comments that improved the
presentation of this work. 
ED acknowledges financial support 
from the project {\it Light-on-Dark} granted by MIUR through
PRIN2017-2017K7REXT. 
MB acknowledges the financial support to this research by INAF, through the Mainstream Grant 1.05.01.86.22 assigned to 
the project ``Chemo-dynamics of globular clusters: the Gaia revolution''.
SK gratefully acknowledges funding from UKRI in the form of a Future Leaders Fellowship (grant no. MR/T022868/1).
EV acknowledges support from NSF grant AST-2009193.

\section*{Data availability}
The raw MUSE data are available from the ESO archive (\url{https://archive.eso.org/eso/eso_archive_main.html}).
Detailed results obtained in this work are available from the authors upon reasonable request.












\end{document}